\documentclass[aps,twocolumn,pra]{revtex4-1}
\usepackage{graphics}
\usepackage{dcolumn}% Align table columns on decimal point
\usepackage{bm}% bold math
\usepackage{amsmath}
\usepackage{hyperref}
\hypersetup{colorlinks=true, urlcolor=blue}
\usepackage{color}% color to the text 
\usepackage{tikz}
\usepackage{multirow}

\usepackage{float}
\usepackage[caption = false]{subfig}
\usepackage{newlfont}
\usepackage{amssymb}
\usepackage{amsfonts}
\usepackage{amsthm}
\usepackage{graphicx}
\usepackage{bm}

\def \be{\begin{equation}}
\def \ee{ \end{equation} }

\begin{document}

\definecolor{red}{rgb}{1,0,0}
\title{Multiparty Quantum Mutual Information: An alternative definition}
%\title{Generalized Multiparty Operational Quantum Mutual Information \\ and Quantum Discord}
%\title{}
\author{Asutosh Kumar}
\email{asutoshk@imsc.res.in}

\affiliation{Harish-Chandra Research Institute, Chhatnag Road, Jhunsi, Allahabad 211 019, India\\
The Institute of Mathematical Sciences, CIT Campus, Taramani, Chennai 600113, India\\
Homi Bhabha National Institute, Anushaktinagar, Mumbai 400094, India}

\begin{abstract}
Mutual information is the reciprocal information that is common to or shared by two or more parties. 
Quantum mutual information for bipartite quantum systems is non-negative, and bears the interpretation of total correlation between the two subsystems. This may, however, no longer be true for three or more party quantum systems. 
In this paper we propose an alternative definition of multipartite information taking into account the shared information between two and more parties. It is non-negative, observes monotonicity under partial trace as well as completely positive maps, and equals multipartite information measure in literature for pure states.
We then define multiparty quantum discord, and give some examples. Interestingly, we observe that quantum discord increases when measurement is performed on a large number of subsystems. Consequently, the symmetric quantum discord, which involves measurement on all parties, reveals the maximal quantumness. This raises question on the interpretation of measured mutual information as classical correlation.

\end{abstract}

\maketitle
  
%\section{Introduction}
{\it Introduction}.--
%%%%%%%%%%%%%%%%%%%%%%
Quantum correlations \cite{horodecki09,modidiscord} are essential ingredients in quantum information theory \cite{nielsen}. 
%Entanglement, in particular, is the characteristic trait of quantum mechanics \cite{schrodinger}. 
Various quantum correlations, different in nature and types, find huge applications in quantum information processing tasks. Consequently, their characterization and quantification is inevitable. Several non-classical correlation measures have been proposed for bipartite quantum systems, and some of them have been extended to multipartite settings. 
Nonetheless, quantifying multipartite quantum correlations in quantum physical systems remains a challenging problem. Recently however, significant developments have been made towards this end in the form of multipartite global (symmetric) quantum discord (GKD) \cite{gqd}, conditional entanglement of multipartite information (CEMI) \cite{cemi}, quantum correlations relativity (QCR) \cite{qcr}. In Ref. \cite{wilde}, an operational interpretation of GQD was given in terms of the partial state distribution protocol. It was also shown that GQD nearly vanishes for a multiparty quantum state that is approximately locally recoverable after performing measurements on each of the subsystems. An important aspect to notice is that all these developments count on some multipartite information measure (see below) \cite{watanabe}.
Multipartite (mutual) information, the reciprocal information that is common to or shared by two or more parties, has an authoritative stand 
%is an indispensable candidate
in the arena. 
Quantum mutual information (QMI), whose definition is motivated by that of classical mutual information (CMI), is well defined for bipartite quantum systems. QMI of a bipartite quantum state \(\rho_{AB}\) is defined as
\begin{align}
I(\rho_{AB})&=S(\rho_{A})+S(\rho_{B})-S(\rho_{AB}) \nonumber \\
&=S(\rho_{AB}\parallel \rho_{A} \otimes \rho_{B}) \geq 0,
\end{align}
where $S(\rho)=-\mbox{tr}(\rho \log \rho)$ is the von Neumann entropy and $S(\rho \parallel \sigma)=\mbox{tr}(\rho \log \rho-\rho \log \sigma)$ is the quantum relative entropy. 
It is non-negative, and bears the interpretation of total correlation between the two subsystems \cite{total-correlation-winter}. It is defined as the amount of work (noise) that is required to erase (destroy) the correlations completely. These properties (non-negativity, interpretation of total correlation) may, however, no longer be true for three or more party quantum systems. The existing quantum version of multipartite information in literature, due to Watanabe \cite{watanabe}, is a straightforward generalization of bipartite QMI
\begin{align}
I_x(A_1:A_2:\cdots:A_n)&=\sum_{k=1}^n S(\rho_{A_K})-S(\rho_{A_1A_2\cdots A_n}) \nonumber \\
&=S(\rho_{A_1A_2 \cdots A_n}\parallel \rho_{A_1}\otimes \cdots \otimes \rho_{A_n}) \nonumber \\
& \geq 0. 
\label{qmi-watanabe}
\end{align}
We refer to it as \emph{conventional} quantum mutual information (CQMI). It is the sum of the individual von Neumann entropies less the joint von Neumann entropy of a multipartite quantum system, $\rho_{A_1A_2\cdots A_n}$. It is non-negative, and monotone non-increasing under the local discarding of information (i.e., $I_x(A_1X_1:A_2X_2:\cdots:A_nX_n) \geq I_x(A_1:A_2:\cdots:A_n)$ for a multiparty quantum state $\rho_{A_1X_1A_2X_2\cdots A_nX_n}$. However, unlike two-party quantum mutual information $I_x(A_1:A_2)$, it does not have any operational interpretation.\\ 
 
In another approach, three-variable CMI \cite{dissension} is defined as
\begin{align}
K(A:B:C)=K(A:B)-K(A,B|C),
\end{align}
where $K(A:B)=H(A)-H(A|B)=H(A)+H(B)-H(A,B)=H(B)-H(B|A)$ is two-variable CMI, $K(A,B|C)=H(A|C)+H(B|C)-H(A,B|C)$ is three-variable conditional mutual information, and $H(.)$ is Shannon entropy. Though both $K(A:B)$ and $K(A,B|C)$ are non-negative, the three-variable CMI can be negative. Using the chain rule $H(X,Y)=H(X)+H(Y|X)$, the following expressions of CMI are equivalent:
\begin{align}
K_1(A:B:C)=[H(A)+H(B)+H(C)] \nonumber \\
-[H(A,B)+H(A,C)+H(B,C)]+H(A,B,C) \\
K_2(A:B:C)=H(A,B)-H(B|A)-H(A|B) \nonumber \\
-H(A|C)-H(B|C)+H(A,B|C) \\ 
K_3(A:B:C)=[H(A)+H(B)+H(C)] \nonumber \\
-[H(A,B)+H(A,C)]+H(A|B,C).
\end{align}
The above definitions of CMI can be extended to the quantum domain. They are obtained by replacing the random variables by density matrices and Shannon entropies by von Neumann entropies, with appropriate measurements in the quantum conditional entropies. Hence,
\begin{eqnarray}
I(A:B:C)=[S(A)+S(B)+S(C)] \nonumber \\
-[S(A,B)+S(A,C)+S(B,C)]+S(A,B,C) \\
J_1(A:B:C)=S(A,B)-S_{{\cal M}}(B|A)-S_{{\cal M}}(A|B) \nonumber \\
-S_{{\cal M}}(A|C)-S_{{\cal M}}(B|C)+S_{{\cal M}}(A,B|C) \\
J_2(A:B:C)=[S(A)+S(B)+S(C)] \nonumber \\
-[S(A,B)+S(A,C)]+S_{{\cal M}}(A|B,C)
\end{eqnarray}
where $S(X) \equiv S(\rho_X)$, and $S_{{\cal M}}(X|Y)=S(\rho_{X|{\cal M}^Y})$ is the quantum conditional entropy obtained after some generalized measurement ${\cal M}$ has been performed on subsystem \(Y\).
It is asserted that the above quantum expressions are not equivalent as measurement assumes its role in the quantum conditional entropies. These QMIs have certain drawbacks. First, surprisingly enough $I(A:B:C)$ is identically zero for arbitrary three-party pure quantum states \cite{dissension} implying that mutual information among the subsystems of three-party pure quantum systems is zero. This is not true in the case of bipartite QMI. Second, $I(A:B:C)$ and other versions of QMI can be negative \cite{dissension,neg-qmi}.
%the three-party QMI $I(A:B:C)=I(A:B)-I(A,B|C)$, interpreted as the measure of the reduction in the uncertainty of one event because of the occurrence of the other event, can be negative. 
How is this negative correlation useful for quantum information tasks? Though the existing definition of three-party QMI is argued to reveal the true nature of quantum correlations \cite{dissension}, the fact that QMI, being a measure of correlation, can assume negative value is challenging.
%its hard to digest why it should take negative value
This perplexing stance, handicapped with any operational interpretation of multipartite information, motivated us to propose an alternative definition of multiparty QMI.\\ 
Our multipartite quantum mutual information for quantum state $\rho_{A_1A_2\cdots A_n}$ assumes the following form
\begin{align}
I(A_1:A_2:\cdots:A_n)&=\sum S(X_{k_1}X_{k_2}\cdots X_{k_{n-1}}) \nonumber \\
&-(n-1)S(A_1A_2\cdots A_n),
\label{qmi-ntotal-def}
\end{align}   
where $X_{k_i} \in \{A_1,A_2,\cdots,A_n\}$.
It takes into account the shared information among \(m\)-parties, $2 \leq m \leq n$, and not only the common information among all parties. This is quite reasonable as information can be distributed or stored among \(m\)-parties. We show that it is non-negative, and argue that it, by its very construction, manifests total correlation. Also it equals CQMI for pure states. Moreover, we obtain its lower and upper bounds in terms of CQMI.\\
 
The rest of this paper is organised as follows: in the following section we provide an alternative definition of multiparty quantum mutual infromation, compute its value for some typical states, and prove its non-negativity and monotonicity. Then, in the next section, we discuss multiparty quantum discord and present a few illustrations. Surprisingly, we observe that the symmetric quantum discord reveals the maximal quantumness. Finally, we conclude.\\  
%In Sec. \ref{op-qmi}, we provide operational definition of multiparty quantum mutual information, and compare with other mutual information. In Sec. \ref{op-qd}, we discuss multiparty quantum discord and present a few illustrations. We show that the symmetric version of quantum discord reveals the maximal quantumness. We finally conclude in Sec. \ref{summary}. 

%\section{Operational Quantum Mutual Information}
%\label{op-qmi}
{\it Quantum Mutual Information}.--
We propose an alternative definition of multiparty quantum mutual information, via the Venn diagram approach, for an \(n\)-party quantum state $\rho_{A_1A_2\cdots A_n}$. As information can be distributed or stored among \(m\)-parties, $2 \leq m \leq n$, our definition takes into account the shared information among \(m\)-parties, and not only the common information among all parties. This can be understood readily using a Venn diagram.\\

\begin{center}
\begin{figure}[ht]
\subfloat[]{\includegraphics[width=1.4in, angle=0]{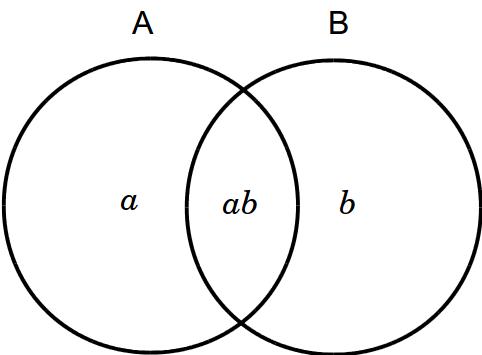}}\hspace{0.2cm}
\subfloat[]{\includegraphics[width=1.4in, angle=0]{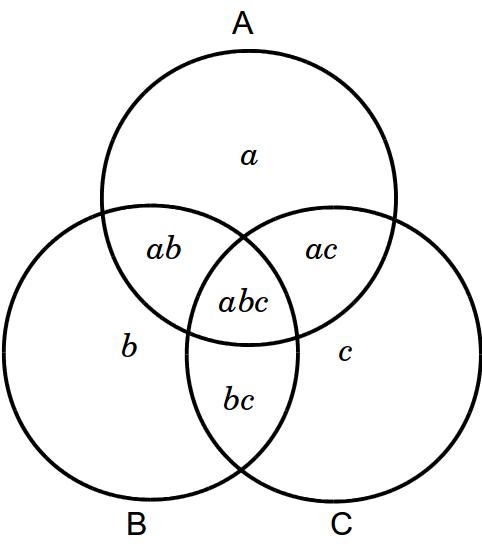}}\\
\subfloat[]{\includegraphics[width=1.4in, angle=0]{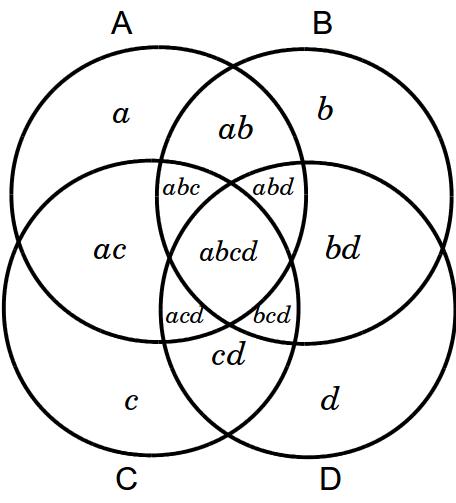}}\hspace{0.2cm}
\subfloat[]{\includegraphics[width=1.4in, angle=0]{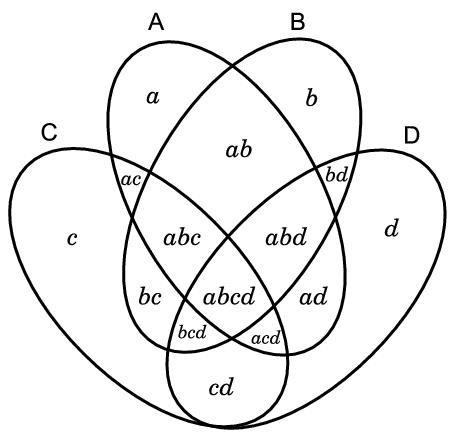}}
\caption{(a) Two-variable, (b) three-variable, and (c \& d) four-variable, Venn diagrams with possible intersecting regions. While (c) does not represent the true Venn diagram of four-variables, (d) does.}
\label{venn234scan}
\end{figure}
\end{center}

%\subsection{Two-party QMI}
{\it Two-party QMI}:
From Fig. \ref{venn234scan}(a), we see that the only way the subsystems \(A\) and \(B\) interact with each other is via region \(ab\), i.e., $ab=A\cap B=A+B-A\cup B$. In entropy language, this translates as $I(A:B)=S(A)+S(B)-S(AB)$. This is usual bipartite QMI.
%\begin{equation}
%I(A:B)=S(A)+S(B)-S(AB).
%\end{equation}

%\subsection{Three-party QMI}
{\it Three-party QMI}:
From Fig. \ref{venn234scan}(b), the possible ways the subsystems \(A\), \(B\) and \(C\) interact among each other are via region \(abc\), which is common to all three, and  regions \(ab\), \(ac\), \(bc\), which are pairwise common. Taking just the region \(abc\), i.e., $abc=A\cap B\cap C=A+B+C-(A\cup B+A\cup C+B\cup C)+A\cup B \cup C$, this ``common information" translates into $I_c(A:B:C):=[abc]=[S(A)+S(B)+S(C)]-[S(AB)+S(AC)+S(BC)]+S(ABC)$.
%\begin{eqnarray}
%I_c(A:B:C):=[abc] \nonumber \\
%=[S(A)+S(B)+S(C)] \nonumber \\
%-[S(AB)+S(AC)+S(BC)]+S(ABC).
%\label{qmi3common}
%\end{eqnarray}
Note that in doing so we have discarded pairwise interactions. However, {\em a priori}, there is no reason to throw them away. Moreover, they can provide important information when examined together. Then ``two-party shared information" reads as $I_{s2}(A:B:C):=[ab+ac+bc]$.
%\begin{equation}
%I_{s2}(A:B:C):=[ab+ac+bc].
%\label{qmi3shared2}
%\end{equation} 
Thus, the total three-party QMI is the sum of the common information and the pairwise shared information: $I(A:B:C)=I_c(A:B:C)+I_{s2}(A:B:C):=[A\cup B\cup C-(a+b+c)]$.
%\begin{align}
%I(A:B:C)&=I_c(A:B:C)+I_{s2}(A:B:C) \nonumber \\
%&:=[A\cup B\cup C-(a+b+c)].  
%\label{qmi3total}
%\end{align}
%Eq. (\ref{qmi3total})  
After simple algebra, $I(A:B:C)$ can be expressed in entropy language as
\begin{align}
I(A:B:C)&=S(AB)+S(AC)+S(BC)-2S(ABC) \nonumber \\
&=S\left(\rho_{ABC}^{\otimes 2}\parallel \rho_{AB}\otimes \rho_{AC}\otimes \rho_{BC}\right).
\end{align} 
%After simple algebra, $I(A:B:C)$ can be expressed in entropy language, thereby in terms of bipartite QMIs, as
%\begin{align}
%I(A:B:C)&=S(AB)+S(AC)+S(BC)-2S(ABC) \nonumber \\
%&=\frac{1}{3}\left[2\sum I(X_1:X_2X_3)-\sum I(X_1:X_2)\right] \nonumber \\
%& \geq 0,
%\end{align} 
%where $X_i \in \{A,B,C\}$. The last inequality follows since $I(X_1:X_2X_3) \geq I(X_1:X_2)$. Thus, by construction, three-party OQMI is non-negative. 
It guarantees that $I(A:B:C)$ is not identically zero for arbitrary three-party pure quantum systems.

%\subsection{Four-party QMI}
{\it Four-party QMI}:
Fig. \ref{venn234scan}(c) does not represent the true Venn diagram of four variables as pairwise interacting regions \(ad\) and \(bc\) are missing. The correct four-variable Venn diagram is represented in Fig. \ref{venn234scan}(d). The total four-party QMI is then defined as $I(A:B:C:D):=[A\cup B\cup C\cup D-(a+b+c+d)]$ which, again,
after some simple algebra, can be expressed as \cite{note1}
\begin{align}
I(A:B:C:D)&=\sum_{X_1,X_2,X_3}S(X_1X_2X_3)-3S(ABCD) \nonumber \\
&=S\left(\rho_{ABCD}^{\otimes 3}\parallel \bigotimes_{\{X_i\}} \rho_{X_1X_2X_3}\right),
\label{qmi4total-ent}
\end{align}
where $X_i \in \{A,B,C,D\}$. 
We list in Table \ref{table-qmi} the values of common information ($I_c$) and QMI ($I$) of some typical states (see also Fig. \ref{qmi3}).

An \(n\)-party QMI can be analogously defined (see Eq.~(\ref{qmi-ntotal-def}))
%\begin{widetext}
%\begin{align}
%%I(A_1:A_2:\cdots:A_n)&=I_c(A_1:A_2:\cdots:A_n) \nonumber \\
%%&+\sum_{k=2}^{n-1} I_{sk}(A_1:A_2:\cdots:A_n) \nonumber \\
%%&:=[A_1A_2\cdots A_n] \nonumber \\
%%&+\sum_{k=2}^{n-1} \left[\sum_{i_1<i_2<\cdots <i_k}A_{i_1}A_{i_2}\cdots A_{i_k}\right] \nonumber \\
%I(A_1:A_2:\cdots:A_n)&:=[A_1\cup A_2\cup \cdots \cup A_n 
%-(a_1+a_2+\cdots+a_n)] \nonumber \\
%&=\sum_{\{X_{k_i}\}} S(X_{k_1}X_{k_2}\cdots X_{k_{n-1}}) 
%-(n-1)S(A_1A_2\cdots A_n) \nonumber \\
%&=S\left(\rho_{A_1A_2\cdots A_n}^{\otimes n-1}\parallel \bigotimes_{\{X_{k_i}\}} \rho_{X_{k_1}X_{k_2}\cdots %X_{k_{n-1}}}\right),
%\label{qmi-ntotal-def}
%\end{align} 
%\end{widetext}
\begin{widetext}
\begin{align}
I(A_1:A_2:\cdots:A_n)&:=[A_1\cup A_2\cup \cdots \cup A_n 
-(a_1+a_2+\cdots+a_n)] \nonumber \\
&=\sum_{k=1}^n S(\rho_{\overline{A_k}}) 
-(n-1)S(\rho_{A_1A_2\cdots A_n}) \nonumber \\
&=\sum_{k=1}^n S\left(\rho_{A_1A_2\cdots A_n}\parallel \rho_{A_k}\otimes \rho_{\overline{A_k}}\right) - S\left(\rho_{A_1A_2\cdots A_n}\parallel \bigotimes_{k=1}^n \rho_{A_k}\right) \nonumber \\ 
&=S\left(\rho_{A_1A_2\cdots A_n}^{\otimes n-1}\parallel \bigotimes_{k=1}^n \rho_{\overline{A_k}}\right),
\label{qmi-ntotal-def}
\end{align} 
\end{widetext}
where \(n\)-party common information is evaluated as $I_c(A_1:A_2:\cdots:A_n)=\sum_{k=1}^n (-1)^{k+1} \sum_{\{A_{I_k}\}} S_{A_{I_k}}$ with $A_{I_k} \equiv A_{i_1}A_{i_2}\cdots A_{i_k}$, $S_X \equiv S(\rho_X)$ and $S_{\overline{A_i}} \equiv S(\rho_{\overline{A_i}}) = S(\rho_{A_1\cdots A_{i-1}A_{i+1}\cdots A_n})$. Hence, QMI can be re-written as $I(A_1:A_2:\cdots:A_n)=\sum_{i=1}^n S_{\overline{A_i}}-(n-1)S_{A_1A_2\cdots A_n}$. {\em An \(n\)-party quantum mutual information is the sum of \((n-1)\)-party von Neumann entropies less \((n-1)\) times the joint von Neumann entropy of an \(n\)-party quantum system.} 
We argue here that multiparty QMI $I(A_1:A_2:\cdots:A_n)$, like $I(A_1:A_2)$,  by very construct, bears the interpretation of total correlation of a multiparty quantum system. 
%like the two-party quantum systems \cite{total-correlation-winter}.
Further, we can obtain generalized QMI by replacing the von Neumann entropy, $S(\rho)=-\mbox{tr}(\rho \log \rho)$, with generalized entropies like the Renyi entropy, $S_q^R(\rho)=\frac{1}{1-q} \log[\mbox{tr}(\rho^q)]$ \cite{renyi-entropy}, the Tsallis entropy, $S_q^T(\rho)=\frac{1}{1-q}[\mbox{tr}(\rho^q)-1]$ \cite{tsallis-entropy}, and the (smooth) min-max entropies \cite{renner-dutta}. Both Renyi and Tsallis entropies reduce to von Neumann entropy in the limit $q\rightarrow 1$.

In subsequent theorems, we prove that $I(A_1:A_2:\cdots:A_n)$ is non-negative, and is monotonically non-increasing under partial trace and completely positive maps. It equals CQMI for pure states, and give its lower and upper bounds in terms of CQMI.\\ 

\textbf{Theorem 1.} \textit{$I(A_1:A_2:\cdots:A_n)$ is non-negative.}\\
%\texttt{Proof.} 
%Let us denote $S_X \equiv S(\rho_X)$. First, we will prove the theorem for three- and four parties, and then generalize our argument to arbitrary \(n\)-party case. 
Non-negativity of quantum mutual information $I(A_1:A_2:\cdots:A_n)$ follows directly from that of quantum relative entropy, $S(\rho \parallel \sigma) \geq 0$. Here we provide an alternative proof.
To prove this, we will extensively use a variant of strong subadditivity relation, $S_{XYZ}+S_Y \leq S_{XY}+S_{YZ}$, which states that conditioning reduces entropy, i.e., $S_{X|YZ} \leq S_{X|Y}$. The proof for \(n\)-party case follows as \cite{note2}:
$I(A_1:A_2:\cdots:A_n)=S_{12\cdots (n-1)}+S_{12\cdots (n-2)n}+\cdots
+S_{23\cdots n}-(n-1)S_{12\cdots n}
=S_{12\cdots (n-1)}-S_{1|23\cdots (n-1)n} 
-S_{2|13\cdots (n-1)n}-\cdots -S_{(n-1)|12\cdots (n-2)n}
 \geq S_{12\cdots (n-1)}-S_{1|23\cdots (n-1)} 
-S_{2|13\cdots (n-1)}-\cdots -S_{(n-1)|12\cdots (n-2)}  
= \cdots 
 \geq S_{123}-S_{1|23}-S_{2|13}-S_{3|12} 
=S_{12}-S_{1|23}-S_{2|13}
 \geq S_{12}-S_{1|2}-S_{2|1} 
=S_1+S_2-S_{12} \geq 0$.
Hence, the theorem is proved.
\hfill $\blacksquare$\\

\textbf{Theorem 2.} \textit{$I(A_1:A_2:\cdots:A_n)$ equals CQMI for pure states.}\\
%\texttt{Proof.}
For the pure quantum states, $S_{A_1A_2\cdots A_n}=0$ and $S_{\bar{A_i}}=S_{A_i}$. Hence, $I(A_1:A_2:\cdots:A_n)=I_x(A_1:A_2:\cdots:A_n)$ using Eq. (\ref{qmi-ntotal-def}) and Eq. (\ref{qmi-watanabe}).
\hfill $\blacksquare$\\

\textbf{Theorem 3.} \textit{$I(A_1:A_2:\cdots:A_n)$ observes monotonicity under partial trace, and any completely positive map $\Phi$.}\\
%\texttt{Proof.}
These are direct consequences of the monotonicity of quantum relative entropy \cite{ruskai} under partial trace, $S(\rho_A \parallel \sigma_A) \leq S(\rho_{AX} \parallel \sigma_{AX})$, and any completely positive map $\Phi$, $S\left(\Phi(\rho) \parallel \Phi(\sigma)\right) \leq S(\rho \parallel \sigma)$. 
\hfill $\blacksquare$\\

%In following theorem, we obtain lower and upper bounds on OQMI in terms of the conventional QMI $I_x\equiv I_x(A_1:A_2:\cdots :A_n)$.\\ 
\textbf{Theorem 4.} \textit{$I_x-(n-2)S_{12\cdots n} \leq I \leq I_x+2S_{12\cdots n}$.}\\
%\texttt{Proof.} 
Using strong subadditivity entropic relation, $S_X+S_Y \leq S_{XZ}+S_{YZ}$, and Araki-Lieb inequality, $S_X-S_Y \leq S_{XY} \Rightarrow S_X-S_{XY} \leq S_{Y}$, we can, respectively, obtain $\sum_{i=1}^n S_{A_i} \leq \sum_{i=1}^n S_{\bar{A_i}}$ and 
$\sum_{i=1}^n S_{\bar{A_i}}-nS_{A_1A_2\cdots A_n} \leq \sum_{i=1}^n S_{A_i}$.
Therefore, $I(A_1:A_2:\cdots :A_n)=\sum S_{A_{k_1}A_{k_2}\cdots A_{k_{n-1}}} 
-(n-1)S_{A_1A_2\cdots A_n} \geq \sum_{i=1}^n S_{A_i}-(n-1)S_{A_1A_2\cdots A_n} 
=I_x(A_1:A_2:\cdots :A_n)-(n-2)S_{A_1A_2\cdots A_n}$. Again, $I(A_1:A_2:\cdots :A_n)=\sum S_{A_{k_1}A_{k_2}\cdots A_{k_{n-1}}}-(n-1)S_{A_1A_2\cdots A_n}
= \big(\sum_{i=1}^n S_{\bar{A_i}}-nS_{A_1A_2\cdots A_n}\big) 
+S_{A_1A_2\cdots A_n} \leq \sum_{i=1}^n S_{A_i} + S_{A_1A_2\cdots A_n}
=I_x(A_1:A_2:\cdots :A_n)+2S_{A_1A_2\cdots A_n}$. Hence, the proof.
\hfill $\blacksquare$\\
The lower bound being dependent on \(n\) is weak. Moreover, we find numerically that for an \(n\)-party quantum system $\rho_{A_1A_2\cdots A_n}~(n=3,4)$, we have
\begin{equation}
0 \leq I^{(n)}-I_x^{(n)} \leq I_x^{(n)} - \sum I^{(2)},
\end{equation}
where $I^{(k)}$ is the \(k\)-party quantum mutual information, and the inequality is saturated for $n=3$ (this can be shown analytically).\\ 
%It suggests that both $I$ and $I_x$ satisfy monogamy inequality \cite{monogamy, mono-rev}.

\textbf{Theorem 5.} \textit{$I(A_1:A_2:\cdots:A_n)$ is additive for product states.}\\
%\texttt{Proof.} 
Additivity of quantum mutual information $I(A_1:A_2:\cdots:A_n)$ follows directly from that of the von Neumann entropy for product states, $S(\rho \otimes \sigma) = S(\rho) + S(\sigma)$. Here we consider a four-party state for the illustration (the proof is identical irrespective of the number of parties and the partition across which $\rho_{A_1A_2 \cdots A_n}$ is product). Let $\rho_{A_1A_2 A_3 A_4} = \rho_{A_1A_2} \otimes \rho_{A_3A_4}$. Then 
\begin{widetext}
\begin{align}
I(A_1:A_2:A_3:A_4) &= S(A_1A_2A_3) + S(A_1A_2A_4) + S(A_1A_3A_4) + S(A_2A_3A_4) - 3 S(A_1A_2A_3A_4) \nonumber \\
&= S(A_1) + S(A_2) + S(A_3) + S(A_4) + 2[S(A_1A_2) + S(A_3A_4)] - 3[S(A_1A_2) + S(A_3A_4)] \nonumber \\
&= [S(A_1) + S(A_2) - S(A_1A_2)]+ [S(A_3) + S(A_4) - S(A_3A_4)] \nonumber \\
&= I(A_1:A_2) + I(A_3:A_4). 
\end{align}
%Hence, the proof.
%\hfill $\blacksquare$\\

\begin{center}
\begin{figure}[htb]
\subfloat[]{\includegraphics[width=2in, angle=0]{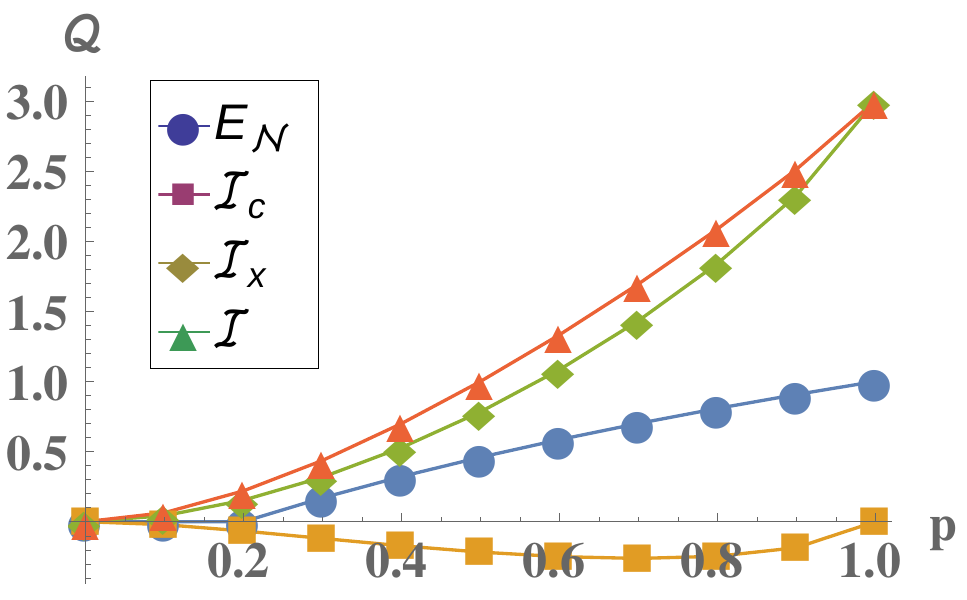}}
\subfloat[]{\includegraphics[width=2in, angle=0]{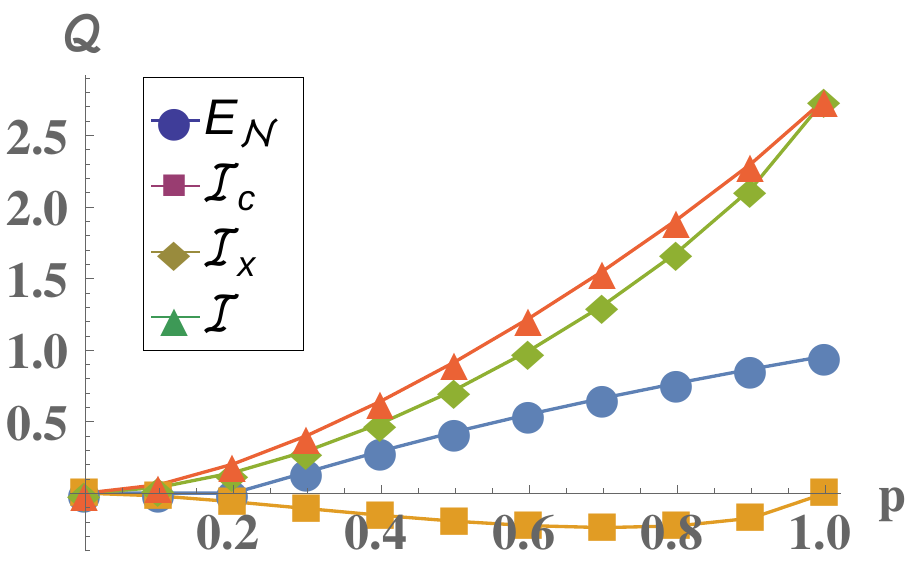}}
\subfloat[]{\includegraphics[width=2in, angle=0]{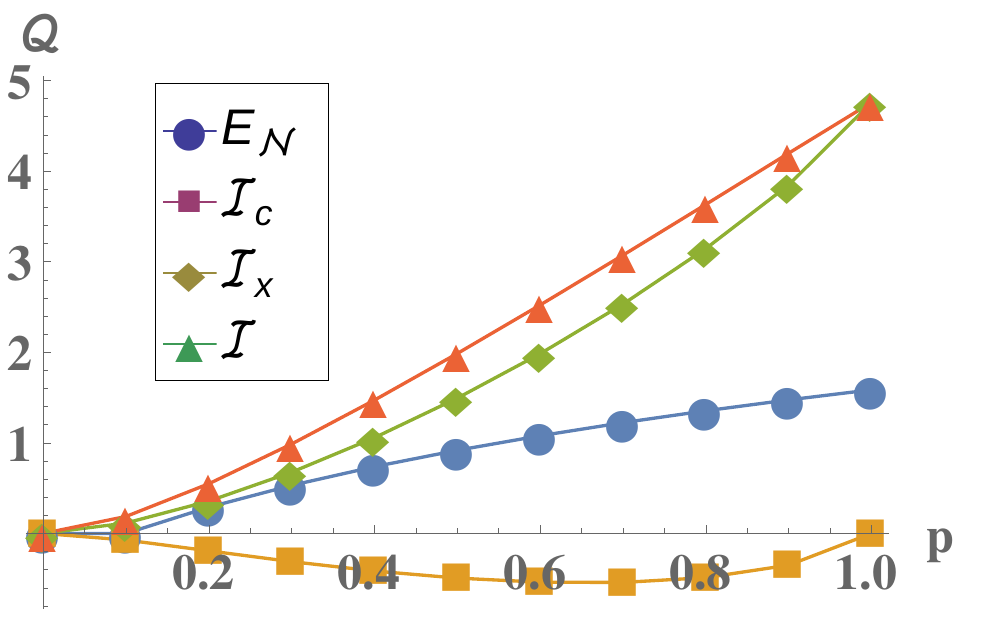}}
\caption{Plots of logarithmic-negativity $E_{\cal N}(A:BC)$ \cite{negativity,logneg}, common information $I_c(A:B:C)$, CQMI $I_x(A:B:C)$, and QMI $I(A:B:C)$ against the white noise parameter \(p\), of three-party state $\rho_{ABC}=p|\psi\rangle \langle \psi|+(1-p)\frac{I}{8}$ for different $|\psi\rangle$s: (a) GHZ state $|GHZ\rangle=
\frac{1}{\sqrt{2}}(|000\rangle+|111\rangle)$ \cite{ghz}, (b) W state $|W\rangle=
\frac{1}{\sqrt{3}}(|001\rangle+|010\rangle+|100\rangle)$ \cite{dicke}, and (c) totally antisymmetric state $|\psi_{as}\rangle=
\frac{1}{\sqrt{6}}(|123\rangle-|132\rangle+|231\rangle-|213\rangle+|312\rangle-|321\rangle)$ \cite{3qutrit-antisym}. 
%,admixed with white noise
In all the three cases, while $I_c(A:B:C)$ vanishes at \(p=0,1\) and is negative at intermediate values, QMI $I(A:B:C)$ vanishes at \(p=0\) only and is positive for other values. QMI is greater than or equal to $I_x(A:B:C)$. We see that logarithmic-negativity, an entanglement measure, exceeds common information indicating that common information cannot be total correlation.}
\label{qmi3}
\end{figure}
\end{center}
\end{widetext}

\begin{table}[htb]
\begin{tabular}{|c|c|c|}
\hline
{\color[HTML]{000000} State} & $I_c$    & $I$      \\ \hline \hline
$|GHZ_2\rangle$             & 2        & 2        \\ \hline
$|GHZ_3\rangle$              & 0        & 3        \\ \hline
$|D_3^1\rangle$              & 0        & 2.75489 \\ \hline
$|\psi_{as}\rangle$          & 0        & 4.75489  \\ \hline
$|GHZ_4\rangle$              & 2        & 4     \\ \hline
$|D_4^1\rangle$              & 0.490225 & 3.24511 \\ \hline
$|D_4^2\rangle$              & 0.490225 & 4 \\ \hline
$|C_4\rangle$                & -2       & 4      \\ \hline
\end{tabular}
\caption{Values of common information ($I_c$) and QMI ($I$) of $|GHZ_n\rangle=\frac{1}{\sqrt{2}}(|0\rangle^{\otimes n}+|1\rangle^{\otimes n})$ \cite{ghz}, $|D_n^r\rangle=\frac{1}{\sqrt{\binom{n}{r}}}\sum_{{\cal P}}{\cal P}[|0\rangle^{\otimes n-r}|1\rangle^{\otimes r}]$ \cite{dicke}, three-qutrit totally antisymmetric state $|\psi_{as}\rangle=
\frac{1}{\sqrt{6}}(|123\rangle-|132\rangle+|231\rangle-|213\rangle+|312\rangle-|321\rangle)$ \cite{3qutrit-antisym}, and four-qubit cluster state $|C_4\rangle=\frac{1}{2}(|0000\rangle+|0011\rangle+|1100\rangle-|1111\rangle)$ \cite{cluster-state}. While $I_c$ can be negative, $I$ is non-negative. For pure states, $I=I_x$, from Theorem 2.}
\label{table-qmi}
\end{table}

%\begin{widetext}
%\begin{center}
%\begin{figure}[htb]
%\subfloat[]{\includegraphics[width=2in, angle=0]{qmi3-ghz1}}
%\subfloat[]{\includegraphics[width=2in, angle=0]{qmi3-w1}}
%\subfloat[]{\includegraphics[width=2in, angle=0]{qmi3-antisym1}}
%\caption{Plots of logarithmic-negativity $E_{\cal N}(A:BC)$ \cite{negativity,logneg}, common information $I_c(A:B:C)$, CQMI $I_x(A:B:C)$, and QMI $I(A:B:C)$ against the white noise parameter \(p\), of three-party state $\rho_{ABC}=p|\psi\rangle \langle \psi|+(1-p)\frac{I}{8}$ for different $|\psi\rangle$s: (a) GHZ state $|GHZ\rangle=
%\frac{1}{\sqrt{2}}(|000\rangle+|111\rangle)$ \cite{ghz}, (b) W state $|W\rangle=
%\frac{1}{\sqrt{3}}(|001\rangle+|010\rangle+|100\rangle)$ \cite{dicke}, and (c) totally antisymmetric state $|\psi_{as}\rangle=
%\frac{1}{\sqrt{6}}(|123\rangle-|132\rangle+|231\rangle-|213\rangle+|312\rangle-|321\rangle)$ \cite{3qutrit-antisym}. 
%%,admixed with white noise
%In all the three cases, while $I_c(A:B:C)$ vanishes at \(p=0,1\) and is negative at intermediate values, QMI $I(A:B:C)$ vanishes at \(p=0\) only and is positive for other values. QMI is greater than or equal to $I_x(A:B:C)$. We see that logarithmic-negativity, an entanglement measure, exceeds common information indicating that common information cannot be total correlation.}
%\label{qmi3}
%\end{figure}
%\end{center}
%\end{widetext} 

%\begin{widetext}
\begin{center}
\begin{figure}[htb]
\subfloat[]{\includegraphics[width=1.5in, angle=0]{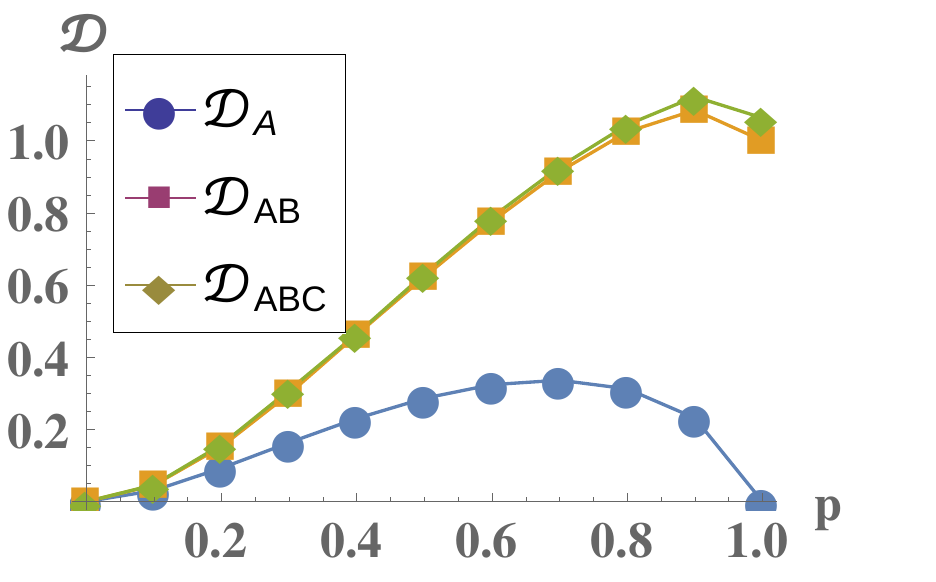}}
\subfloat[]{\includegraphics[width=1.5in, angle=0]{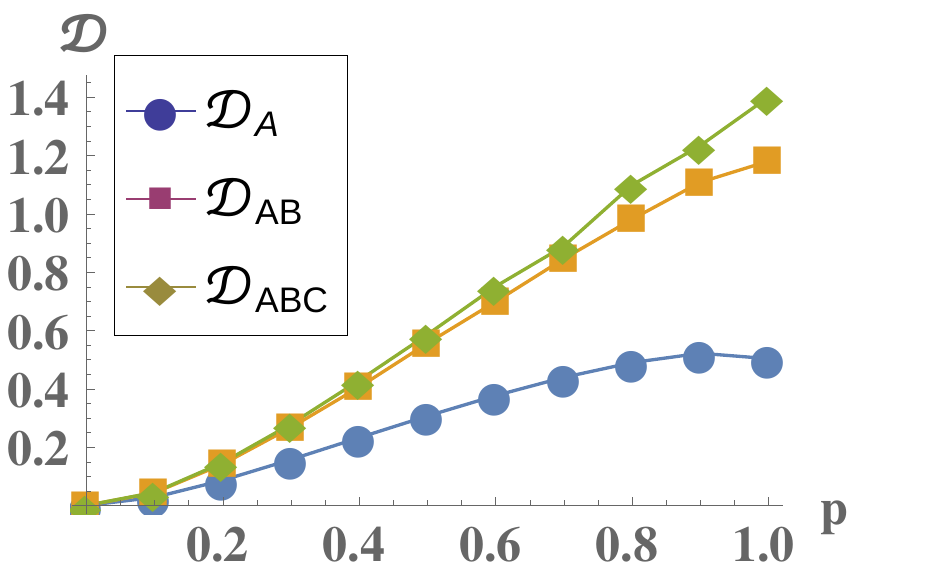}}\\
\subfloat[]{\includegraphics[width=1.5in, angle=0]{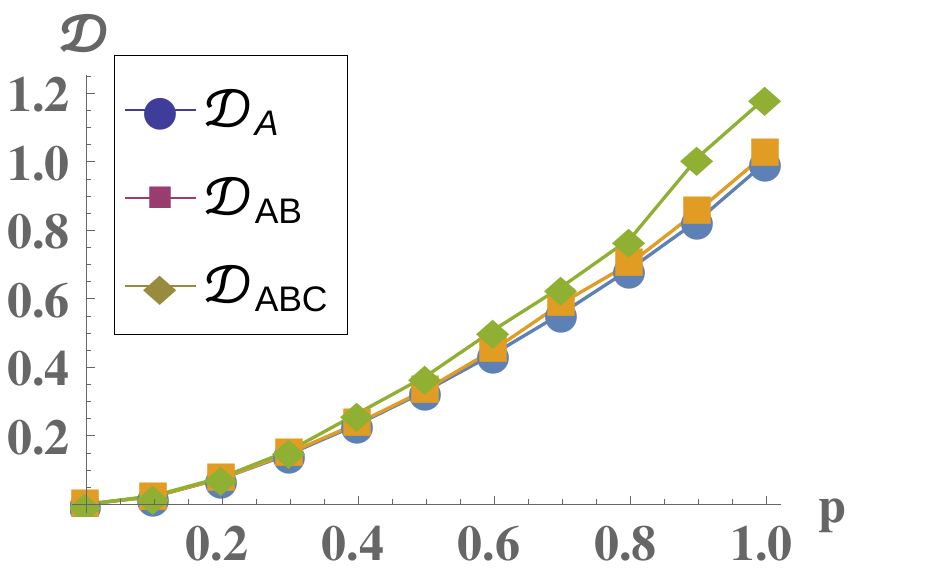}}
\subfloat[]{\includegraphics[width=1.5in, angle=0]{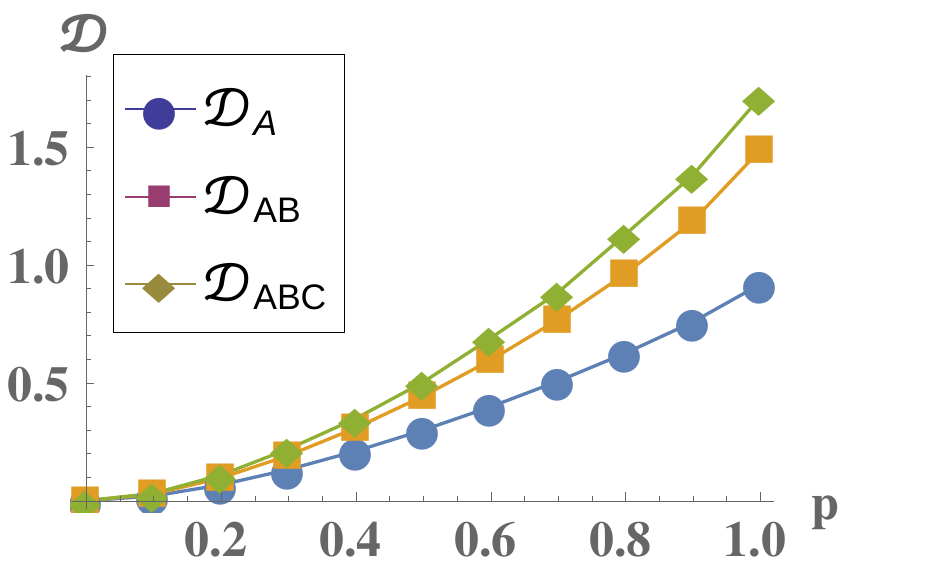}}
\caption{Plots of quantum discords based on alternative definition of QMI (top panel) and \emph{conventional} quantum discords (bottom panel) ${\cal D}_A$, ${\cal D}_{AB}$ and ${\cal D}_{ABC}$ against the white noise parameter \(p\), of three-party states
$\rho_{ABC}=p|GHZ\rangle \langle GHZ|+(1-p)\frac{I}{8}$ (a \& c), and $\rho_{ABC}=p|W\rangle \langle W|+(1-p)\frac{I}{8}$ (b \& d).
%$|GHZ\rangle=\frac{1}{\sqrt{2}}(|000\rangle+|111\rangle)$ (a \& c), and W state $|W\rangle=\frac{1}{\sqrt{3}}(|001\rangle+|010\rangle+|100\rangle)$ (b \& d).  
%,admixed with white noise
Contrary to our intuition, quantum discord increases when measurement is performed on larger number of subsystems. The higher values of quantum discord suggest that W state is more robust, as compared to GHZ state, against measurement. 
%While operational QD is concave, conventional QD is convex.
}
\label{dis-ghz-w}
\end{figure}
\end{center}
%\end{widetext}

%\section{Multiparty Quantum Discord}
%\label{op-qd}
{\it Multiparty Quantum Discord}.--
In this section, we extend the definition of bipartite quantum discord \cite{hv,oz} to multipartite setting. Quantum discord for a bipartite quantum state \(\rho_{AB}\) is defined as ${\cal D}(\rho_{AB})= I(\rho_{AB}) - \max_{{\cal M}} J(\rho_{AB})$, where $I(\rho_{AB})= I(A:B)= S(A)+ S(B)- S(AB)$ and $J(\rho_{AB}) = S(B) - S_{{\cal M}}(B|A)$.
%\begin{equation}
%\label{eq:2qd}
%{\cal D}(\rho_{AB})= I(\rho_{AB}) - \max_{{\cal M}} J(\rho_{AB}),
%\end{equation}
%where 
%\begin{equation}
%\label{qmi}
%I(\rho_{AB})= I(A:B)= S(A)+ S(B)- S(AB),
%\end{equation}
%and
%\begin{equation}
%\label{eq:classical}
%J(\rho_{AB}) = S(B) - S_{{\cal M}}(B|A).
%\end{equation}
Here measurement is performed on subsystem \(A\) with a rank-one projection-valued measurement \(\{A_i\}\), producing the states  
\(\rho_{B|i} = \frac{1}{p_i} \mbox{tr}_A[(A_i \otimes I_B) \rho (A_i \otimes I_B)]\), 
with probability \(p_i = \mbox{tr}_{AB}[(A_i \otimes I_B) \rho (A_i \otimes I_B)]\).
\(I\) is the identity operator on the Hilbert space of \(B\). Hence, the conditional entropy of \(\rho_{AB}\) is given by $ S_{{\cal M}}(B|A) = \sum_i p_i S(\rho_{B|i})$.
%\begin{equation}
% S_{{\cal M}}(B|A) = \sum_i p_i S(\rho_{B|i}).
%\end{equation}

Three-party quantum discord can then be defined, when measurement is performed on subsystem \(A\), subsystem \(AB\) and the whole system, as follows
\begin{eqnarray}
{\cal D}_A(\rho_{ABC})&=& I(\rho_{ABC}) - \max_{\Phi_{A}} I(\Phi_A(\rho_{ABC})) \\
{\cal D}_{AB}(\rho_{ABC})&=& I(\rho_{ABC}) - \max_{\Phi_{AB}} I(\Phi_{AB}(\rho_{ABC}))
\end{eqnarray}
and 
\begin{eqnarray}
{\cal D}_{ABC}(\rho_{ABC})&=& I(\rho_{ABC}) - \max_{\Phi_{ABC}} I(\Phi_{ABC}(\rho_{ABC}))
\label{qd-sym}
\end{eqnarray}
where $I(\sigma_{XYZ})=I(X:Y:Z)$, $\Phi_A(\rho_{ABC})=\sum_i \Phi_{A_i} \rho_{ABC} \Phi_{A_i}$, $\Phi_{AB}(\rho_{ABC})=\sum_{i,j} \Phi_{A_iB_j} \rho_{ABC} \Phi_{A_iB_j}$, and $\Phi_{ABC}(\rho_{ABC})=\sum_{i,j,k} \Phi_{A_iB_jC_k} \rho_{ABC} \Phi_{A_iB_jC_k}$
%\begin{eqnarray}
%I(\sigma_{XYZ})&=&I(X:Y:Z) \nonumber \\
%\Phi_A(\rho_{ABC})&=&\sum_i \Phi_{A_i} \rho_{ABC} \Phi_{A_i} \\
%\Phi_{AB}(\rho_{ABC})&=&\sum_{i,j} \Phi_{A_iB_j} \rho_{ABC} \Phi_{A_iB_j} \\
%\Phi_{ABC}(\rho_{ABC})&=&\sum_{i,j,k} \Phi_{A_iB_jC_k} \rho_{ABC} \Phi_{A_iB_jC_k} 
%\end{eqnarray}
with $\Phi_{A_i}=\pi_i \otimes I \otimes I$, $\Phi_{A_iB_j}=\pi_i \otimes \pi_j \otimes I$, and $\Phi_{A_iB_jC_k}=\pi_i \otimes \pi_j \otimes \pi_k$. 
Eq. (\ref{qd-sym}) is the symmetric quantum discord or global quantum discord (GQD) \cite{gqd}.
%which represents the loss of correlation due to measurement. 
Similarly, multiparty quantum discord can be defined.

%\section{Symmetric Quantum Discord is the Maximal}
%\label{sym-qd}
Quantum discord, employing von Neumann entropy, of three-party GHZ state and W state admixed with white noise is shown in Fig. \ref{dis-ghz-w}. Quite unexpectedly, we find that ${\cal D}_A \leq {\cal D}_{AB} \leq {\cal D}_{ABC}$, that is, quantumness increases when measurement is performed on a large number of subsystems. This observation seems to be independent of the definition of quantum mutual information. The symmetric quantum discord, which requires measurement on all the parties, reveals the maximal quantumness.
%Plausibly, this is because when we perform measurement on lesser number of subsystems, we throw away classical correlation as well as some amount of quantum correlation.
This contradicts the interpretation of measured mutual information as classical information because measuring more than one subsystem should yield more classical information and hence less quantum discord.

%\section{Conclusion}
%\label{summary}
{\it Conclusion}.--
To sum up, we have proposed an alternative definition of quantum mutual information for multipartite setting. It is non-negative, and obeys monotonicity under partial trace and any completely positive map. We argue that it manifests total correlation of a multiparty quantum system. We then employed this definition of quantum mutual information to define multiparty quantum discord. Surprisingly, we found that more quantumness can be harnessed by performing measurement on larger number of parties which is quite counter-intuitive. The symmetric quantum discord reveals the maximal quantumness. This suggests that measured mutual information should not be interpreted as classical correlation. We believe that our work will provide further insights in understanding the nature of non-classical correlations.\\ 
%In particular, if the QMI can be associated with multiparty channel capacity. The bipartite CMI is the largest possible or limiting rate of reliable communication \cite{thomas-cover}.
 
%\begin{acknowledgements}
A.K. thanks Ujjwal Sen for useful suggestions and illuminating discussions, and acknowledges research fellowship from the Department of Atomic Energy, Government of India.
%AK acknowledges computations performed at cluster computing facilities at HRI.
%\end{acknowledgements}

%%%%%%%%%%%%%%%%%%%%%

\begin{thebibliography}{99}

\bibitem{horodecki09} R. Horodecki, P. Horodecki, M. Horodecki, and K. Horodecki, Rev. Mod. Phys. {\bf 81}, 865 (2009).

\bibitem{modidiscord} K. Modi, A. Brodutch, H. Cable, T. Patrek, and V. Vedral, Rev. Mod. Phys. {\bf 84}, 1655 (2012).

\bibitem{nielsen} M.A. Nielsen and I.L. Chuang, {\em Quantum Computation and Quantum Information} (Cambridge University Press, Cambridge, 2000).

\bibitem{gqd} M. Piani, P. Horodecki, and R. Horodecki, Phys. Rev. Lett. {\bf  100}, 090502 (2009); C. C. Rulli and M. S. Sarandy, Phys. Rev. A {\bf 84}, 042109 (2011).

\bibitem{cemi} D. Yang, M. Horodecki, and Z. D. Wang, Phys. Rev. Lett. {\bf 101}, 140501 (2008).

\bibitem{qcr} M. Dugic, M. Arsenijevic, and J. Jeknic-Dugic, Sci China-Phys Mech Astron {\bf 56(4)}, 732 (2013).

\bibitem{wilde} M. M. Wilde, Proceedings of the Royal Society A {\bf 471}, 20140941 (2015).

\bibitem{watanabe} S. Watanabe, IBM Journal of Research and Development {\bf 4(1)}, 66 (1960).

\bibitem{total-correlation-winter} B. Groisman, S. Popescu, and A. Winter, Phys. Rev. A \textbf{72}, 032317 (2005).

%\bibitem{thomas-cover} T. Cover and J. Thomas, {\em Elements of Information Theory} (John Wiley \& Sons, 1991).

\bibitem{dissension} I. Chakrabarty, P. Agrawal, and A. K. Pati,  Eur. Phys. J. D {\bf 65}, 605 (2011).

\bibitem{neg-qmi} M. Horodecki, J. Oppenheim, and A. Winter, Nature {\bf 436}, 673 (2005).

%\bibitem{schrodinger} E. Schr\"{o}dinger, ``\emph{Discussion of probability relations between separated systems}", Proceedings of the Cambridge Philosophical Society \textbf{31}, 555 (1935).

%\bibitem{teleportation} C. H. Bennett, G. Brassard, C. Cr\'{e}peau, R. Jozsa, A. Peres, and W. K. Wootters, Phys. Rev. Lett. \textbf{70}, 1895 (1993).

%\bibitem{densecoding} C. H. Bennett and S. J. Wiesner, Phys. Rev. Lett. \textbf{69}, 2881 (1992).

%\bibitem{asu-bell-like} A. Kumar, arXiv:1404.6206 [quant-ph].

%\bibitem{asu-mono} A. Kumar, R. Prabhu, A. Sen(De), and U. Sen, arXiv:1312.6640 [quant-ph].

\bibitem{note1} This expression for quantum mutual information can also be obtained using Fig. \ref{venn234scan}(c) even though it does not represent the true Venn diagram of four variables.

\bibitem{ghz} D. M. Greenberger, M. A. Horne, and A. Zeilinger, in {\it Bell’s Theorem, Quantum Theory, and Conceptions of the Universe}, edited by M. Kafatos (Kluwer Academic, Dordrecht, 1989).

\bibitem{dicke} R. Dicke, Phys. Rev. {\bf 93}, 99 (1954).

\bibitem{3qutrit-antisym} I. Jex, G. Alber, S. M. Barnett, and A. Delgado, Fortschr. Phys. {\bf 51}, 172 (2003).

\bibitem{cluster-state} H. J. Briegel and R. Raussendorf, Phy. Rev. Lett. {\bf 86}, 910 (2001); R. Raussendorf and H. J. Briegel, Phy. Rev. Lett. {\bf 86}, 5188 (2001).

\bibitem{renyi-entropy} A. Renyi, in Proc. Fourth Berkeley Symp. Math. Stat. Prob. {\bf 1}, 547 (1961).

\bibitem{tsallis-entropy} C. Tsallis, J. Stat. Phys. {\bf 52}, 479 (1988).

\bibitem{renner-dutta} R. Renner, {\em Security of quantum key distribution}, PhD
thesis, ETH Zurich, 2005, arXiv:quant-ph/0512258; N. Dutta and R. Renner,  IEEE Transactions on Information Theory {\bf 55}, 2807 (2009).

\bibitem{note2} The proof of non-negativity of quantum mutual information for three-party case goes as: $I(A_1:A_2:A_3)=S_{12}+S_{13}+S_{23}-2S_{123}=S_{12}-S_{1|23}-S_{2|13}
\geq S_{12}-S_{1|2}-S_{2|1}=S_1-S_{1|2}=S_1+S_2-S_{12} \geq 0.$

\bibitem{negativity} G. Vidal and R. F. Werner, Phys. Rev. A \textbf{65}, 032314 (2002).

%\bibitem{fan-pra75} Y.-C. Ou and H. Fan, Phys. Rev. A \textbf{75}, 062308 (2007).

\bibitem{logneg} M. B. Plenio, Phys. Rev. Lett. {\bf 95}, 090503 (2005).

\bibitem{ruskai} M. B. Ruskai, J. Math. Phys. {\bf 43}, 4358 (2002).

%\bibitem{monogamy} V. Coffman, J. Kundu, and W. K. Wootters, Phys. Rev. A {\bf 61}, 052306 (2000); T. Osborne and F. Verstraete, Phys. Rev. Lett. {\bf 96}, 220503 (2006).

%\bibitem{mono-rev} H. S. Dhar, A. K. Pal, D. Rakshit, A. Sen(De), and U. Sen, arXiv: 1610.01069 [quant-ph].

\bibitem{hv}L. Henderson and V. Vedral, J. Phys. A {\bf 34}, 6899 (2001).

\bibitem{oz} H. Ollivier and W.H. Zurek, Phys. Rev. Lett. {\bf 88}, 017901 (2001).


%\bibitem{concurrence} S. Hill and W.K. Wootters, Phys. Rev. Lett. {\bf 78}, 5022 (1997); W.K. Wootters, Phys. Rev. Lett. {\bf 80}, 2245 (1998).

%\bibitem{eof} C. H. Bennett, D. P. DiVincenzo, J. A. Smolin, and W. K. Wootters, Phys. Rev. A, \textbf{54}, 3824 (1996).

%\bibitem{ggm} A. Sen(De) and U. Sen, Phys. Rev. A {\bf 81}, 012308 (2010); A. Sen(De) and U. Sen, arXiv:1002.1253 [quant-ph]; R. Prabhu, S. Pradhan, A. Sen(De), and U. Sen, Phys. Rev. A {\bf 84}, 042334 (2011).

%\bibitem{gm} A. Shimony, Ann. N. Y. Acad. Sci. {\bf 755}, 675 (1995); H. Barnum and N. Linden, J. Phys. A {\bf 34}, 6787 (2001); T.-C. Wei and P. M. Goldbart, Phys. Rev. A {\bf 68}, 042307 (2003). 

\end{thebibliography}
\end{document}